\documentclass[aps,pre,showpacs,showkeys,twocolumn,groupedaddress,floatfix]{revtex4-1}
\usepackage{graphicx}
\usepackage{hyperref}
\usepackage{amsthm,amsmath,amssymb}  % AMS

\begin{document}

\title{Two-step electrical percolation in nematic liquid crystal filled by multiwalled carbon nanotubes}
\author{Serhiy Tomylko}
\email{tomulkosv@ukr.net}
\affiliation{Institute of Physics, NAS of Ukraine, Prospect Nauky 46,  03028, Kyiv, Ukraine}
\author{Oleg Yaroshchuk}
\affiliation{Institute of Physics, NAS of Ukraine, Prospect Nauky 46,  03028, Kyiv, ~Ukraine}
\email[Correspondence author:]{o.yaroshchuk@gmail.com}
\author{Nikolai Lebovka}
\email{lebovka@gmail.com}
\affiliation{Institute of Biocolloidal Chemistry named after F.D. Ovcharenko, NAS of Ukraine, Prospect Vernadskogo 42,  03142,  Kyiv, Ukraine}
\date{\today}

\begin{abstract}
Percolation of carbon nanotubes (CNTs) in liquid crystals (LCs) opens way for a unique class of anisotropic hybrid materials with a complex dielectric
constant widely controlled by CNT concentration. Percolation in such systems is commonly described as a one-step process starting at a very low loading of CNTs.
In the present study the two-step percolation was observed in the samples of thickness 250 $\mu$m obtained by pressing the suspension between two substrates.
The percolation concentrations for the first and second percolation processes were $C_n^{p_1}\approx  0.0002$ wt. \% and   $C_n^{p_2}\approx 0.5$ wt. \%, respectively.
The two-stage nature of percolation was explained on a base of mean field theory assuming core-shell structure of CNTs.
\end{abstract}


%Nematic liquid crystals structure of, 61.30.-v
%Nanotubes carbon, 61.48.De
%Percolation in phase transitions, 64.60.ah
%Composite materials electrical conductivity, 72.80.Tm
%Liquid crystals dielectric properties of, 77.84.Nh

\maketitle

\section{Introduction}
Carbon nanotubes (CNTs) have extraordinarily large shape anisotropy (the length-to-diameter ratio may be as large as 500-1000), high mechanical strength, large anisotropy of electric and thermal conductivity, and high chemical resistance. Different types of electrochemical biosensors, strain sensors, electromagnetic switches, screens and other multifunctional devices  made from materials containing CNTs were already proposed
~\cite{Brady-Estevez2010,Endo2008,Hasan2009,Pandey2012,Valentini2013}.

Nowadays, liquid crystals (LC) filled by CNTs attract both scientific and practical interest
~\cite{Zakri2007,Trushkevych2008,Dolgov2010,Basu2010,Scalia2010}. The orientational ordering of LC host imposes ordering of CNTs and has impact on their aggregation~\cite{Dierking2004,Schoot2008}. Moreover, ordering direction of CNTs can be easily controlled by external electric or magnetic field~\cite{Dierking2004,Schoot2008}. On the other hand, doping of LC by CNTs may essentially improve electro-optic performance of LC cells. It allows for reduction of the response time and driving voltage, as well as suppression of undesirable back flow and image retention~\cite{Rahman2009}. Also, the nanotubes bring unusual properties to LCs, such as much enhanced permittivity and electrical conductivity~\cite{Yaroshchuk2014} as well as remarkable electro-optical~\cite{Dolgov2010} and electromechanical~\cite{Basu2008} memory effects.

In the majority of previous works, the concentration of CNTs in LC suspension was rather small ($\leq  10^{-3}$ wt. \%) in order to avoid their essential aggregation.  At the same time, increasing of CNT concentration enhances properties of these particles in the composites and leads to a number of exciting features, such as percolation phenomena and accompanying memory effect ~\cite{Dolgov2010,Dolgov2008,Dolgov2012,Dolgov2009,Yaroshchuk2010}.

The percolation phenomena in the colloidal systems based on CNTs are caused by three-dimensional continuous networks formed by these particles in the dispersion media at some concentration of CNTs called percolation concentration or percolation point, $C_n^p$. Usually, this structural process is accompanied by abrupt increase of electrical conductivity~\cite{Bauhofer2009} and mechanical rigidity~\cite{Sahimi1998} in the vicinity of $C_n^p$. The value of   $C_n^p$ can be controlled by many different factors, such as distribution of quality of electrical contacts/junctions between different particles~\cite{Calberg1999}, presence of interfacial shells around particles embedded in the continuous matrix~\cite{Sushko2013}, clustering, agglomeration~\cite{Aguilar2010} or segregation of particles~\cite{Zhang1998}, variation in the shape and orientation of aggregates and local particle concentration~\cite{Nettelblad2003}, dependence of the network structure upon the particle concentration~\cite{Kovacs2009} and  orientational ordering of CNTs inside a matrix~\cite{Haggenmueller2000,Behnam2007a,Behnam2007,Dombovari2010}.  As for the last, alignment of CNTs can essentially affect percolation characteristics. The high ordering of CNTs destroys the percolation pathways created by intersected nanotubes and thus decreases electrical conductivity. However, Monte Carlo simulations indicate that maximal conductivity can be achieved for slightly aligned rather than isotropically distributed CNTs~\cite{Du2005}. This is in good agreement with the increase of electrical conductivity in CNT composites caused by magnetic field~\cite{Choi2003} or mechanical shear~\cite{Lanticse2006}.

Fundamentally new opportunities for studying the influence of orientational ordering of CNTs on percolation characteristics are opened when using LC matrices. In contrast to isotropic polymer matrices in which the orientational order is formed under external action like extrusion or shear, this order in LCs occurs spontaneously due to molecular self-assembling. The structure of LC mesophases is rather sensitive to external factors such as heat, electric or magnetic field. Thus ordering of CNTs integrated in LC can easily be tuned by soft acting on the LC host.

Despite such opportunities, the experimental data on percolation phenomena in the LC-CNTs systems are quite scarce~\cite{Dolgov2010,Kovalchuk2008,Goncharuk2009,Lebovka2008}. They entirely refer to percolation of electric conductivity in thin cells ($d<20 \mu$m) and can be reduced to the following conclusions:
 \begin{enumerate}
   \item The percolation in LCs is essentially a single-step process with rather low percolation point (    $C_n^p\approx  0.01$ wt. \%).
   \item The percolation concentration  depends upon the phase state of LC medium. In particular, in nematic phase of 5CB, the percolation point was found at  $C_n^p\approx  0.01$ wt. \%, whereas it noticeably increased up to 0.1 wt. \% in isotropic phase ~\cite{Dolgov2010}.
   \item By reaching the percolation transition, the dominant mechanism of electrical conductivity becomes a charge tunneling and hopping between single nanotubes. The next strengthening of percolation network results in domination of mechanism typical for nanotube bundles. These changes in the character of conductivity are explained by improvement of electrical contacts between the percolating nanotubes ~\cite{Dolgov2010}.
 \end{enumerate}

These properties are in good agreement with the properties of the percolation transition of CNTs in polymer matrices. The exception makes only the first conclusion consisting in a single-stage nature of the transition. The matter is that these systems demonstrate fuzzy type or multiple transitions with two or even more percolation thresholds reflecting different stages of formation of percolation network ~\cite{Kovacs2007,Kovacs2009,Zhang1998}.

In the present paper, to clarify the nature of the percolation transition in the system LC-CNTs, we analyze more thoroughly the dependence of the electrical conductivity $\sigma$  on the concentration of nanotubes $C_n$ using different methods of filling of the composites in LC cells. This research helped us to find optimal method of filling which naturally leads to two-stage percolation typical for the dispersions of CNTs in other matrices. In the following, we analyze this percolation it terms of scaling law, present model of this process and its simplest mathematical description.

The rest of the paper is constructed as follows. In Section~\ref{sec:experimental} we describe the materials, technical details used for preparation of samples and methods. Sections~\ref{sec:results} and \ref{sec:discussion} present our main findings. In Section~\ref{sec:conclusion}, we summarize the results and conclude the paper.

\begin{figure*}
  \centering
  \includegraphics[clip=on,width=\textwidth]{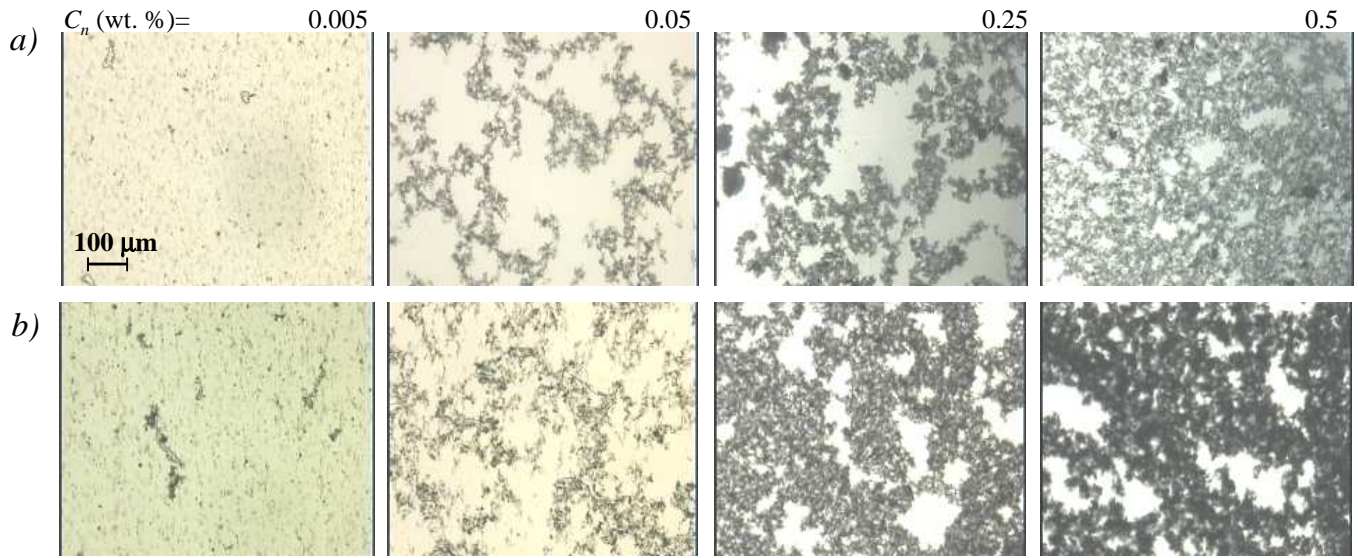} \\
  \caption{ Microphotographs of the layers of 5CB-CNT suspensions filled in the cells with a thickness of 50 $\mu$m. Concentration of CNTs in the samples $1$, $2$, $3$ and $4$ is $0.005$, $0.05$, $0.25$ and $0.5$ wt. \%, respectively.  Series (a) and (b) correspond, respectively, to C- and P- series of samples. It is evident that the amount of nanotubes in both series is comparable up to $C_n = 0.05$ wt. \%, but becomes different at higher $C_n$. The data correspond to nematic phase, $T=25^o$ C. \label{fig:Microphotographs}}
\end{figure*}

\section{Experimental\label{sec:experimental}}
\subsection{Materials\label{sec:materials}}
The nematic liquid crystal 5CB (Merck) with the nematic-to-isotropic transition at $35.5^o$ C and the crystal-to-nematic transition at $22.5^o$C was used in this study. The entangled multi-walled carbon nanotubes (CNTs) were prepared from ethylene using the chemical vapor deposition method (TMSpetsmash Ltd., Kyiv, Ukraine) with Fe-Al-Mo catalyst~\cite{Melezhik2005}. CNTs were further purified by alkaline and acidic solutions and washed by distilled water until reaching the distilled water pH and conductivity values in the filtrate. The typical outer diameter of CNTs, estimated from electrical microscopy images, was $ \approx 30$ nm, while their length ranged from $10$ to $20$ $\mu$m ~\cite{Lisetski2011}. The specific surface area of the CNT powder, determined by N$_2$ adsorption, was $130\pm 5$ m$^2$/g. The specific electric conductivity $\sigma_n$ of the powder of CNTs compressed at 15 TPa was about 10 S/cm along the axis of compression.

\subsection{Preparation of samples}
LCs filled by CNTs were obtained by adding appropriate weights of CNTs ($C_n =0.025-2.0$ wt. \%) to 5CB at $Ò=60^o$C with subsequent $10$ min sonication using an ultrasonic disperser at $22$ kHz and $250$ W. Then composites were incubated at room temperature for $24$ h, sonicated again for $2$ min and loaded into the cells. The volume fraction of CNTs in the composites $\varphi_n$ was estimated as
\begin{equation}\label{eq:varphi}
\varphi_n = (1 + (1/C_n -1)\rho_n/\rho_0)^{-1}\approx C_n \rho_0/\rho_n \approx 0.5 C_n,
\end{equation}
where  $\rho_0$ and  $\rho_n$ are the densities of 5CB and CNTs, respectively. The value of  $\rho_0$ used in this calculation was 1020 kg/m$^3$ ~\cite{Deschamps2008}. The density of the CNTs was assumed to be the same as the density of pure graphite,  $\rho_n = 2045$ kg/m$^3$.

Two methods were used for loading the composites into the cell:
\begin{enumerate}
  \item Filling the composites by capillary forces in the preassembled cells (C-cell).
  \item Pressing of small amount of the composite between two substrates forming a cell (P-cell).
\end{enumerate}

The cells were made from two glass substrates, containing patterned ITO electrodes and layers of polyimide AL3046 (JSR, Japan) designed for planar alignment. The spin coated polyimide films were properly backed and rubbed by a fleecy cloth in order to provide a uniform planar alignment of LC in the field-off state. The cells were assembled so that the rubbing directions of the opposite aligning layers were antiparallel. The cell gap $d$ was maintained by $50 \mu$m and $250 \mu$m Teflon strips. Cell bonding was performed with an epoxy glue.

\subsection{Methods}
The macroscopic alignment in the cells was tested using a light box and two sheet polarizers, while the microstructure was studied using optical polarization microscope Polam L-213M equipped by digital camera conjugated with a personal computer.

The dielectric studies were conducted by oscilloscopic method. The resistance and capacitance of the LC cells were experimentally measured in a wide frequency range, $f=10^{-1}-10^6$ Hz, and used for calculation of real $\epsilon^{'}$ and imaginary $\epsilon^{''}$ parts of complex dielectric constant  $\epsilon^*=\epsilon^{'}-i\epsilon^{''}$.
The frequency $f=2$ kHz from the range $10< f<10^4$ Hz free of any relaxation processes was selected for further evaluation of permittivity $\epsilon^{'}$   and conductivity $\sigma$  values corresponding to the bulk part of a sample. The AC conductivity  $\sigma$  was estimated from the formula $\sigma=2\pi \epsilon_0 \epsilon^{"}f$ where  $\epsilon_0$ is the electric permittivity of free space.

\subsection{Statistical analysis}
Each measurement was repeated at least three times for calculation of the mean values and root-mean-square errors. The error bars in all figures correspond to the confidence level of 95 \%.

\section{Results\label{sec:results}}
 Figure~\ref{fig:Microphotographs} compares microphotographs of the $50 \mu$m layers of 5CB-CNT suspensions inserted into the C-cells (a) and P-cells (b) with different concentrations of CNTs. It can be seen that the microphotographs are rather similar for C- and P-cells up to the concentration $C_n\approx 0.05$ wt. \%. Above $C_n\approx 0.05-0.1$ wt. \%, the method of cell preparation becomes quite important. The aggregates in P-cells are more compacted as compare with those in corresponding C-cells. This fact can be explained by smaller actual concentration of CNTs inside the C-cells. Indeed, the concentration of CNTs inside the P-cells should be equal to the concentration of CNTs in a bulk suspension, $C_n$. However, the situation in the C-cells is different. When the thickness of the cells $d$ is comparable or smaller than the size of aggregates, only the aggregates with the size smaller than $d$ or individual CNTs are effectively involved by LC subjected to capillary forces in the filling process. It results in selective sampling of CNTs at the edges of the C-cell. The filtered out big aggregates remained at the entrance to the cell can be easily observed in optical microscope.
\begin{figure}%[htbp]
\centering
\includegraphics[width=0.9\linewidth,clip=true]{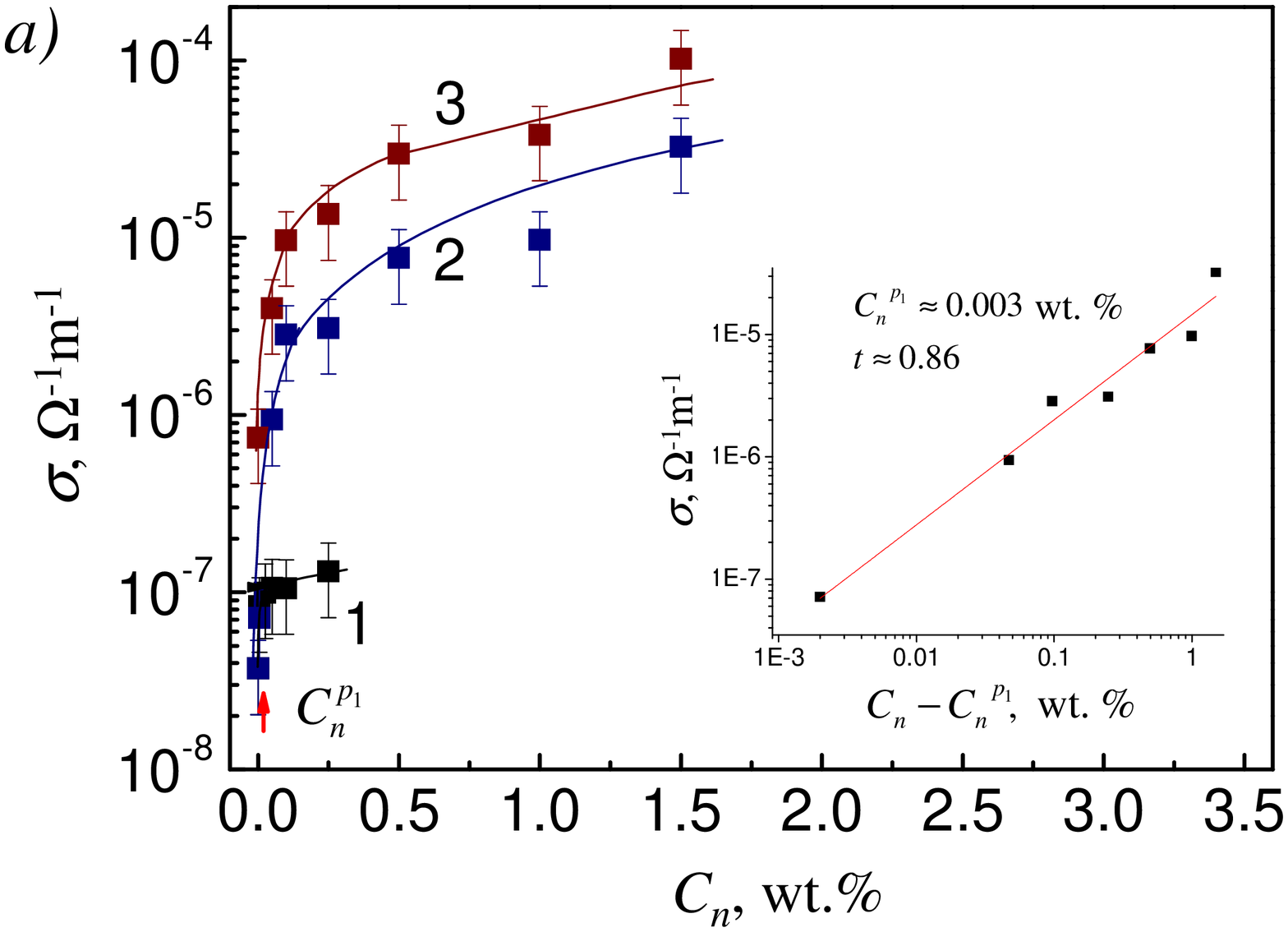} \\
%\hfill
\includegraphics[width=0.9\linewidth,clip=true]{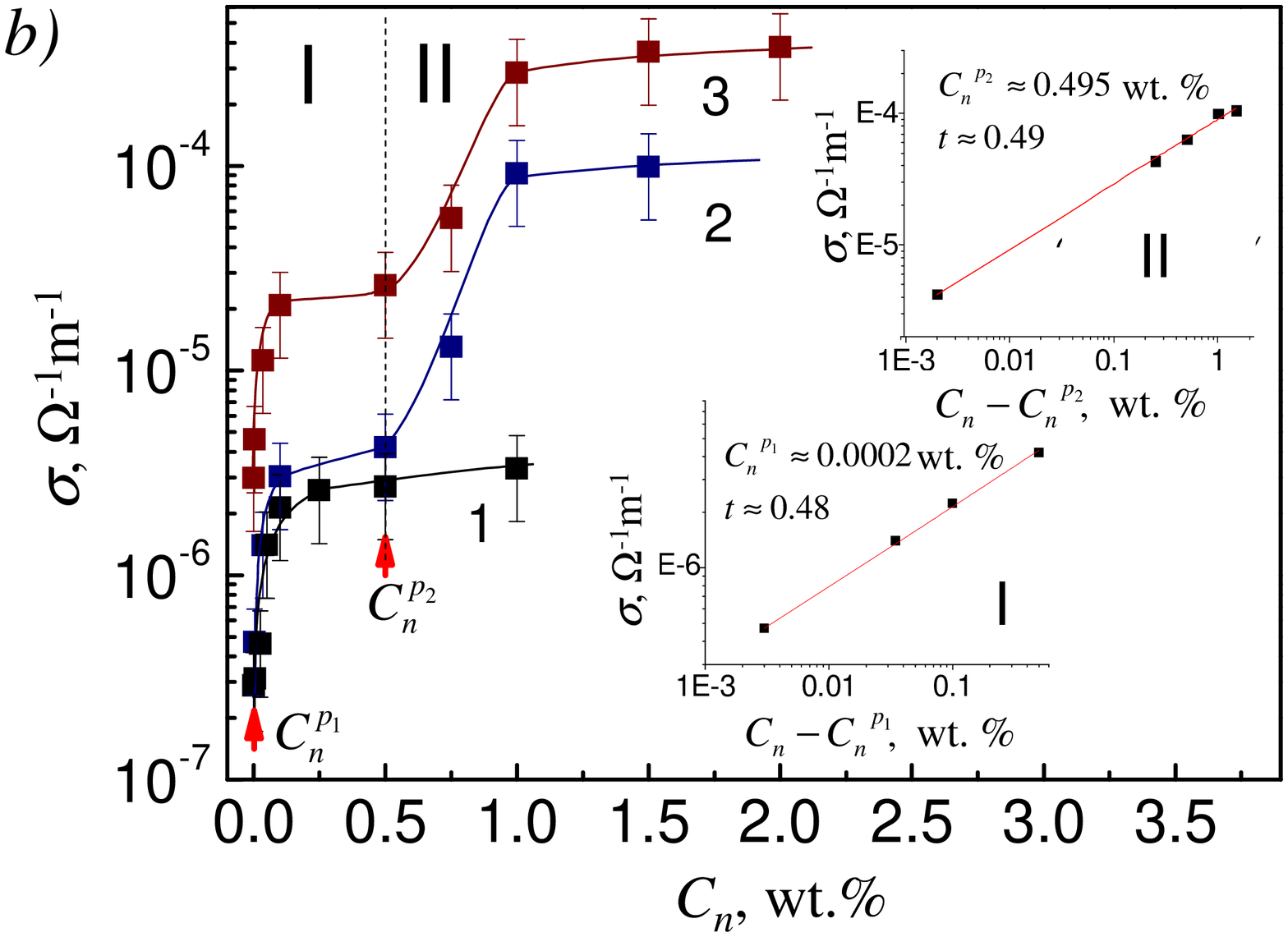}\\
\caption{ Electrical conductivity $\sigma$   versus CNT concentration $C_n$  curves for the layers of 5CB-CNT suspensions with a thickness of $50 \mu$m (a)  and $250 \mu$m (b). Lines are added for guiding eyes. Curve 1 corresponds to C- series, while curves 2 and 3 to P-series. Temperature of measurements was $25^o$ C for curves 1 and 2 and $55^o$C for curve 3.  The numbers I and II in (b) mark concentration ranges of the first and second percolation processes. The insets present curve 2 (P-series, $25^o$C) or its parts in double logarithmic scale along with the fitting curves according to Eq. \ref{eq:perc}.\label{fig:Cond}}
\end{figure}

Concentration dependencies of the electrical conductivity of 5CB-CNT composites, $\sigma(C_n)$, in the C- and P-cells are shown in Fig.~\ref{fig:Cond}. Figures ~\ref{fig:Cond}a and ~\ref{fig:Cond}b correspond to thickness of the cells $50\mu$m and $250\mu$m, respectively. Initially, these curves sharply rise and subsequently demonstrate tendency to saturation. The data evidence that the curves $\sigma(C_n)$ for C- and P- series diverge with increasing of CNT concentration starting at $C_n\approx 0.05$ wt. \% for $d=50 \mu$m and $C_n \approx  0.2$ wt. \% for $d=250 \mu$m, so that the measured electrical conductivity in the C-cells becomes noticeably smaller than in the P-cells. This indicates smaller actual concentration of CNTs inside the C-cell, which is in full correspondence with the data of microphotographs presented in Fig.~\ref{fig:Microphotographs}. Thus, since the actual concentration of CNTs in the C-cells cannot be estimated correctly, the capillary filling is not a suitable method for LC suspensions with enhanced loading of CNTs. Because of this, only the curves corresponding to P-series of samples will further be analyzed.

\begin{figure}%[htbp]
  \centering
  \includegraphics[width=0.9\linewidth]{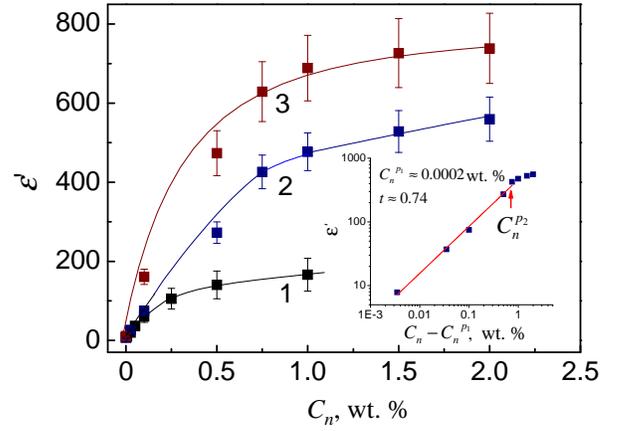}\\
  \caption{Effective dielectric constant of 5CB-CNT suspension $\epsilon^{'}$ as a function of CNT concentration $C_n$  for the samples of thickness $250 \mu$m. Lines are added just for eye guidance.  Curve 1 corresponds to C-series, while curves 2 and 3 to P-series. Temperature of measurements was $25^o$C for the curves 1 and 2 and $55^o$C for the curve 3. The inset represents the curve 2 in double logarithmic scale along with the best fit to it with a function  $\epsilon^{'}\propto (C_n-C_n^{p_1})$. \label{fig:Perm}}
\end{figure}

 Figure ~\ref{fig:Cond} shows distinct difference in the  $\sigma(C_n)$ curves obtained for $d=50 \mu$m and $d=250 \mu$m series of P-cells. The  $\sigma(C_n)$ curve for $d=50 \mu$m demonstrates a single stage growth as in a number of previous studies ~\cite{Lebovka2008,Goncharuk2009,Dolgov2010}. In turn, the curve for $d=250 \mu$m series shows a two-stage percolation character similarly to many polymer dispersions of CNTs ~\cite{Kovacs2007,Kovacs2009,Pang2012}. Finally note that this multistage character is not visible for the concentration dependence of effective permittivity. Figure ~\ref{fig:Perm} shows that the  $\epsilon^{'}(C_n)$ curves monotonically grow with a tendency to saturation and have a smooth shape. The dependencies of  $\sigma(C_n)$ and  $\epsilon^{'}(C_n)$ obtained for isotropic phase ($T=55^o$C) (Figs. ~\ref{fig:Cond} and ~\ref{fig:Perm}) lie above the corresponding curves for nematic phase ($T=25^o$C). This tendency is natural since  $\sigma$ and  $\epsilon^{'}$ in LC grow with temperature. Qualitatively, character of percolation in isotropic phase is the same as in nematic phase. Thus, impact of the phase state on the observed anomalies in percolation was not essential.

\section{Discussion\label{sec:discussion}}
We turn first to the sample preparation procedure for the composites LC-CNTs. In the previous studies of these materials the procedure developed for liquid crystals was used. It consisted in that the LC was filled up in a thin planar cell by means of capillary forces. As shown above, this procedure is acceptable for low concentration of CNTs, but becomes unreliable with increasing of the concentration. The problem is caused by blocking of bigger aggregates of CNTs at entrance to the cell, which intensifies with a growth of CNT concentration. In is naturally that the smaller is the thickness of the cell, the lower is the critical concentration of CNTs at which efficiency of the method of capillary filling is lost (Fig. ~\ref{fig:Cond}). The situation aggravates with lengthening and entanglement of nanotubes that promotes their aggregation. On the contrary, the limit of applicability of capillary filling can be significantly increased by using short-length nanotubes. We demonstrated this by using multiwalled CNTs from Cheap Tubes, USA, having a length of $0.5-2 \mu$m. In this case, critical concentration for capillary filling in $50 \mu$m cells was increased to $0.5$ wt. \%. The details of these studies will be separately published elsewhere.

After determining conditions for observation of the two-stage electrical percolation, move on to the specific of this process. According to Fig.~\ref{fig:Cond}b, the $\sigma(C_n)$ curve demonstrates a sequence of two sharp increases and saturations. This two-stage character seems to be a common feature of fluid composites with low viscosity. In particular, it was previously observed for CNT dispersions in polymer matrices. The first percolation process was attributed to dynamic (kinetic) percolation appearing due to movement and interaction of CNTs. It can be described by dynamic theory ~\cite{Kovacs2007,Bauhofer2009}. A rather low value of threshold concentration observed for this process ($C_n^{p_1} =10^{-3}-10^{-1}$ wt. \% ~\cite{Bauhofer2009,Sandler2003}) was explained by low viscosity of fluid-like matrix, promoting intensive movement of CNTs and, as result, their flocculation. Same mechanism may explain low threshold percolation in LC systems, which was a single percolation detected in previous studies and the first percolation in the present research.

Earlier, by working with thinner samples, we believed that a low value of threshold concentration of CNTs in LCs ($10^{-3}-10^{-2}$ wt. \%) is caused by the fact that the length of CNTs is comparable with the thickness of dispersion layer. This means that even single nanotubes or their small linear aggregates are capable to "short-circuit" the samples causing seeming percolation. However, the results of present work demonstrate that threshold concentration of this process does not depend on cell thickness considerably and so the mentioned explanation is not longer convincing. On the contrary, dynamic nature of network formation explains the low percolation threshold very logically, because nematic LC is a fluid of low viscosity.

The second percolation process in polymer dispersions of CNTs ~\cite{Kovacs2007} was assigned to static percolation developing at higher loading when Brownian motions of CNTs are restricted. It can be described by statistical theories assuming random particle distribution and absence of their movement and interaction. The  $\sigma(C_n)$ curve in this process is usually well fitted to percolation scaling law
\begin{equation}\label{eq:perc}
\sigma\propto (C_n-C_n^p)^t,
\end{equation}
where  $C_n^p$ and $t$ are percolation concentration and transport exponent, respectively. In ~\cite{Kovacs2007}, the second percolation point of  $\sigma(C_n)$ curve was observed as a crossover from saturation state achieved after the first percolation to power law behavior described by Eq.\ref{eq:perc}.

In our case, behavior of  $\sigma(C_n)$ curve is a bit different. Two parts of this curve corresponding to different stages of percolation are qualitatively similar. They consist of rapid growing and saturation sections. The second saturation which was not detected in polymer dispersions of CNTs is probably due to wider concentration range used in our research. Another feature is that both percolation processes are fitted well to scaling law described by Eq.\ref{eq:perc}. This may indicate that dynamic processes play an important role only during formation of CNT network. At the same time, the network formed even at the first percolation stage is sufficiently stable and can be described in frame of statistical theories. The fitting parameters in the first and the second process are  $C_n^{p_1}=0.0002$ wt. \%, $t_1=0.48$ and  $C_n^{p_2}=0.495$ wt. \%, $t_2=0.49$, respectively. Note, that the first concentration approaches zero, because of low viscosity of LC. The transport index $t$ for both transitions is lower than theoretical values for 3d and even 2d percolation ($t=2$ and $t=4/3$, respectively ~\cite{Stauffer1992}. As discussed in
~\cite{Sandler2003}, this is due to strong aggregation of CNTs rather than reduction in system dimensionality. In other words, an actual network is not a true statistical percolation derived by ignoring effective interaction of CNTs.

Note that the  $\sigma(C_n)$ curve measured in isotropic phase fits to the scaling law described by Eq.\ref{eq:perc} too. Moreover, within experimental error, the values
of $C_n^{p_1}$  and $C_n^{p_2}$  are not distinguishable from those of a nematic phase. At the same time, this curve is characterized by smaller transport indices: $t_1=0.37$ and $t_2=0.19$. This indicates a slowing of percolation process, which might be caused by destructive action of Brownian motions of nanotubes and lack of their orientational order.

It may seem strange that the two percolations are not distinguishable in case of series with a cell gap of $50 \mu$m. The  $\sigma(C_n)$ curve for this series can be satisfactorily fitted to a single curve (Eq.\ref{eq:perc}) with the parameters  $C_n^p=0.003$ wt. \% and $t=0.86$. We believe that in this case two percolation processes strongly overlap so that they manifest itself as a single hybrid percolation. A possible reason is that the size of aggregates becomes comparable with a cell gap at much lower concentrations than in case of $250 \mu$m cells. This means that aggregates start to be compressed at lower concentration.
The facts that a threshold concentration  became between $C_n^{p_1}$ and $C_n^{p_2}$  and a scaling exponent $t$ is practically doubled comparing with $t_1$ and $t_2$ in a two-step percolation confirm the assumption on imposing of two percolation processes.

The concentration dependence of the effective permittivity also shows interesting features. As shown in ~\cite{Yaroshchuk2014}, these curve for thin cells ($d<50 \mu$m) can be satisfactorily fitted to scaling law described by Eq.\ref{eq:perc}. However, for the $250 \mu$m series we observed bend of the straight line $\epsilon^{'}(C_n)$ obtained in a double logarithmic scale at a concentration roughly corresponding to  $C_n^{p_2}$. This means that the effective permittivity of the composite is also sensitive to the second percolation of conductivity. We attribute this to increase in polarizability of percolating network with improving of electrical contacts between the nanotubes.

Finally we propose mathematical description for the observed two-step percolation based on various types of contacts between the particles. Previous researches in this direction were devoted exclusively to the polymers filled by CNTs. They assumed distinction of the types of contacts between conducting particles
~\cite{Nettelblad2003,Sheng1982}, distribution of contact resistances for CNTs with different diameters ~\cite{Kovacs2007}, local variations in particle concentration and/or shape and orientation ~\cite{McQueen2004}.

Recently, the possibility of two-step percolation behavior for spherical particles was predicted accounting for the core-shell structure of conductive particles on the base of Bruggeman's effective medium approximation
~\cite{Sushko2013}. The further consideration is based on this approach. It is assumed that the observed two-step percolation in 5CB-CNT dispersions is caused by the core-shell structure of particles with highly conductive CNTs as cores and less conductive LC interfacial layers as shells. Then the percolation thresholds $C_n^{p_1}$ and $C_n^{p_2}$  are associated with percolations through the shells and cores, respectively, and therefore will be further called    $C_n^{s}$ and $C_n^{c}$.

For the sake of simplicity, CNTs will be further considered as quasi-spherical tortuous coils that is quite reasonable for the entangled tubes ~\cite{Lee2013,Lisetski2014}. For the conductive spherical particles with a core-shell structure Bruggeman's equation for electrical conductivity of composite can be represented as ~\cite{Sushko2013}:
\begin{equation}\label{eq:Bruggeman}
\sum_{i=0}^{2}\varphi_i\frac{\sigma_i-\sigma}{\sigma_i+A\sigma},
\end{equation}
where  $\varphi_i$ and $\sigma_i$ are volume fraction and electrical conductivity, respectively, and $A$ is an adjustable parameter determined by the value of percolation threshold and shape of the particle ~\cite{Cai2006}.

\begin{figure}%[htbp]
  \centering
  \includegraphics[width=0.95\linewidth,clip=true]{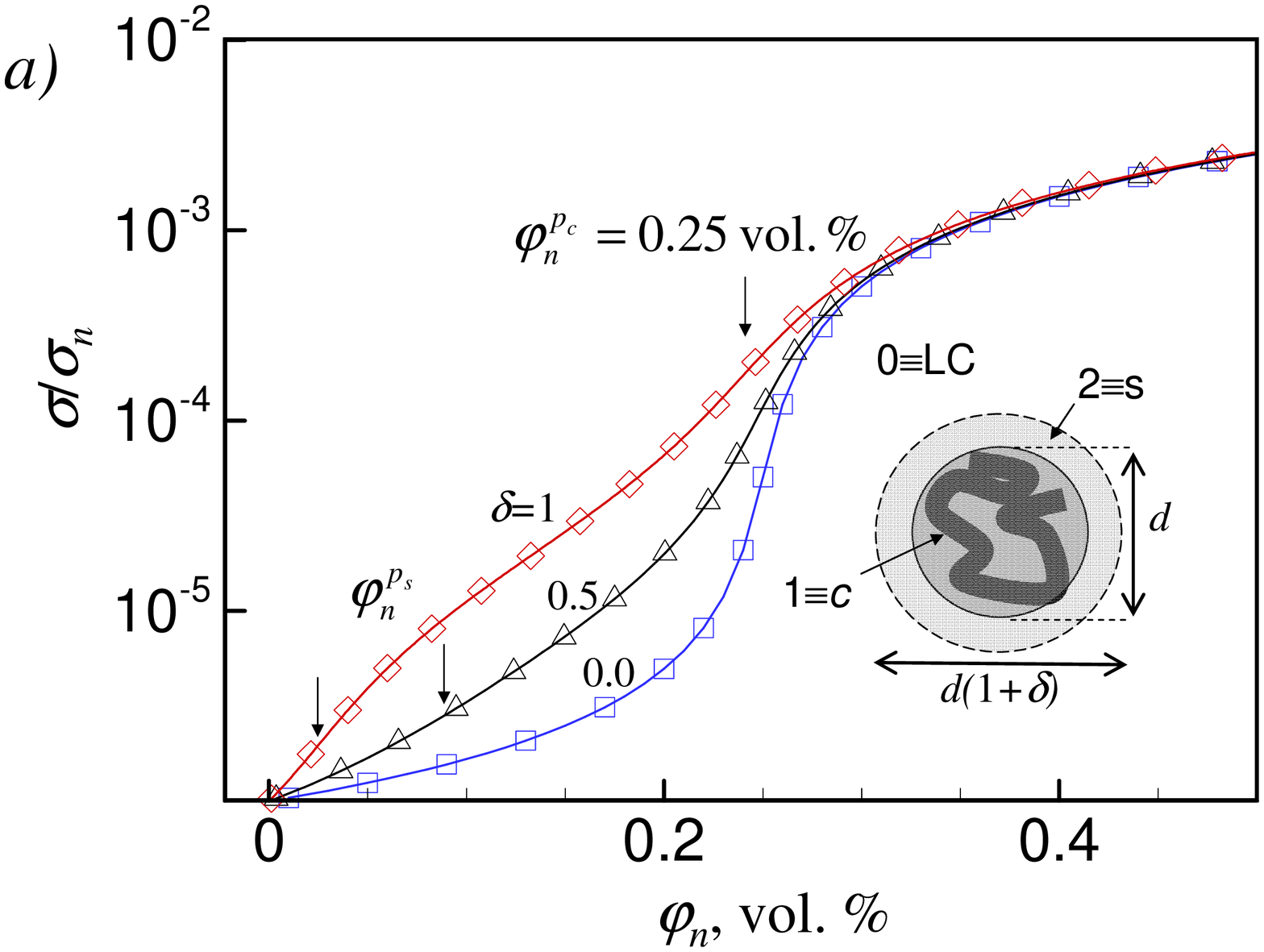}\\
  \includegraphics[width=0.95\linewidth,clip=true]{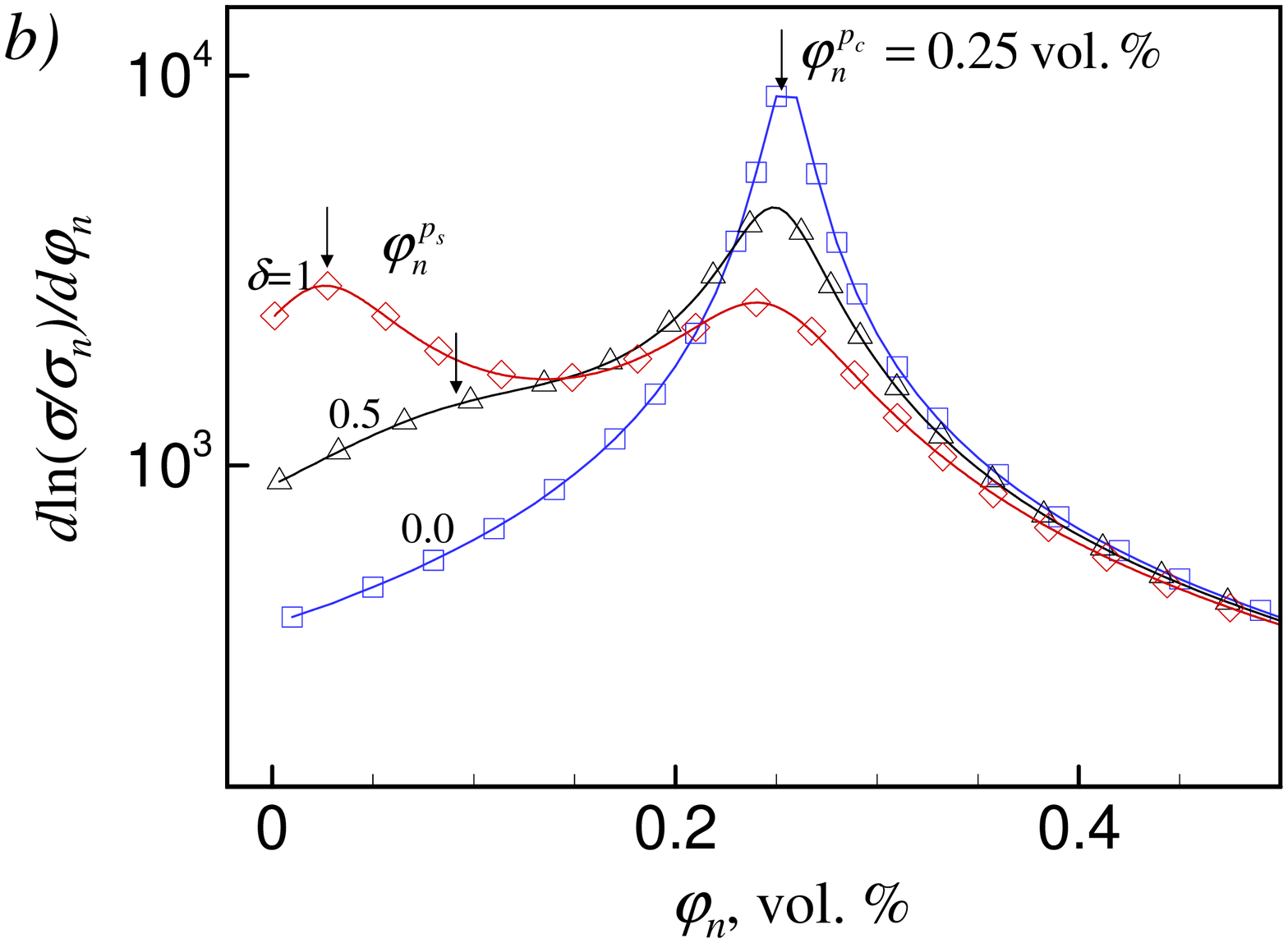}\\
  \caption{Relative electrical conductivity  $\sigma/\sigma_n$ (a) and its derivative $d\ln(\sigma/\sigma_n)/d\varphi_n$ (b)  versus volume fraction of CNTs  $\varphi_n$ calculated for different values of parameter $\delta$  at  $\varphi_n^{p_c}=0.25$ vol. \% ($\approx 0.5$ wt. \%). The calculation was performed assuming that $\sigma_0/\sigma_n=10^{-6}$ and  $\sigma_s/\sigma_n=10^{-3}$.  The inset to (a) presents model of CNT coil with the core-shell structure.\label{fig:Model}}
\end{figure}

Index $i$ for continuous medium is $0$ ( $\sigma_0$,$\varphi_0$), for particle core is $1$ ( $\sigma_1= \sigma_n$, $\varphi_1 = \varphi_n$) and for particle shell is $2$ ($\sigma_2 = \sigma_s$, $\varphi_2=\varphi_s$), (Fig.~\ref{fig:Model}a).

For compact spherical conductive inclusions analyzed in ~\cite{Sushko2013}, $A$ is 2 and percolation threshold  $\varphi_1^p$ is $1/3$. Note that modern phenomenological theory accounts also for the different critical exponents below and above the percolation threshold and estimates $A$ as $(1- \varphi_1^p)/ \varphi_1^p$
~\cite{McLachlan2007}. Numerous experimental data evidenced that for CNT filled composites the value of  $\varphi_1^p$  may be extremely low that may be explained by the giant aspect ratio of CNTs ~\cite{Bauhofer2009}.

Dependence of  $\sigma(\varphi)$ was obtained by solving Eq.\ref{eq:Bruggeman} numerically accounting to $\Sigma\varphi_i=1$, $\varphi_s = \varphi_n(1+\delta)^3$ and  $\varphi_0=1- \varphi_n- \varphi_n(1+\delta)^3$. Here, $\delta$ is the relative enlargement of particle diameter  $\Delta d/d$ due to the shell around the CNT, $d$ is the effective diameter of CNT coil (Fig.~\ref{fig:Model}a). It was also taken into account that $A=(1-\varphi_n^{p_c})/\varphi_n^{p_c}$, where $\varphi_n^{p_c}$ is the volume fraction of CNTs for percolation through the cores of the nanotubes ~\cite{McLachlan2007}.

Figure ~\ref{fig:Model}a presents examples of the calculated  $\sigma/\sigma_n(\varphi_n)$) curves for different values of $\delta$ and  $\varphi_n^{p_c}=0.25$ vol. \% ($\approx 0.5$ wt. \%). The calculation were done using  $\varphi_0/\varphi_n=10^{-6}$,  $\sigma_s/\sigma_n=10^{-3}$. For the given value of percolation threshold through cores $\varphi_n^{p_c}$, the percolation threshold through shells  may be identified as the maximum of the derivative $d\ln(\sigma/\sigma_n)/d\varphi_n$ (Fig. ~\ref{fig:Model}b).

\begin{figure}%[htbp]
  \centering
  \includegraphics[width=0.9\linewidth,clip=true]{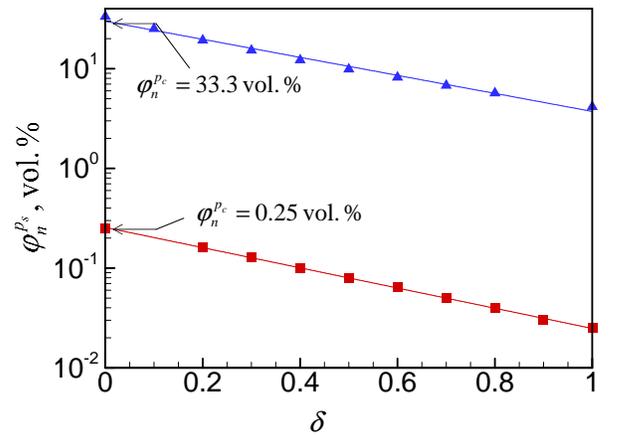}
  \caption{Exemplary dependencies of percolation concentration through shells, $\varphi_n^{p_s}$, versus relative width of the shell $\delta$ calculated for two different values of percolation concentration through cores, $\varphi_n^{p_c} =0.25$ vol. \% and  $\varphi_n^{p_c} =33.3$ vol. \%. The calculations were performed assuming $\sigma_0/\sigma_n=10^{-6}$ and  $\sigma_2/\sigma_n=10^{-3}$.\label{fig:Ratio}}
\end{figure}

The observed two steps reflect two sharp transitions of conductivity $\sigma_0\rightarrow\sigma_s$ and  $\sigma_s\rightarrow\sigma_ñ$. Figure ~\ref{fig:Ratio} presents examples of dependencies of percolation concentration of the first percolation threshold at $\varphi_n=\varphi_n^{p_s}$ (percolation through shells) versus relative width of the shell $\delta$ at two different values of volume fraction $\varphi_n^{p_s}$  corresponding to percolation through cores.

The obtained data show that for thick shells ($\delta\approx 1$) the first percolation threshold, $\varphi_n^{p_s}$, can be noticeably smaller than the second one, $\varphi_n^{p_s}$. It was in correspondence with our experimental observation.

%\clearpage

\section{Conclusion\label{sec:conclusion}}
The electrical conductivity and dielectric constant of CNTs dispersions in nematic LC 5CB has been studied  in dependence of CNT loading,  cell thickness and filling technique. The filling was based on capillary forces (C-cells) or pressing the drop of dispersion between two substrates forming the cell (P-cells). It was demonstrated that at high concentration of CNTs ($C_n>0.5$ wt \%) the typical for LCs technique of capillary filling becomes ineffective for the LC dispersions of entangled CNTs. This happens when the thickness of the cell becomes comparable with the size of bigger aggregates so that only small aggregates or individual CNTs can be effectively involved by capillary forces inside the C-cell. It results in selective sampling of CNTs at the edges of C-cell and lowering of actual concentration of CNTs in the cells comparing with that in the bulk, $C_n$. In this case, the true value of $C_n$ in the cells provides only the pressing method. Understanding of this made it possible to measure correctly the concentration dependences of conductivity and effective permittivity of the dispersions in the wide range $C_n=0-2$ wt. \%. Corresponding dependence  $\sigma(C_n)$ for the series of P-samples with a thickness of $250 \mu$m has clearly shown a two-stage percolation earlier detected only in polymer dispersions of CNTs ~\cite{Kovacs2007}.  Both percolation steps were fitted well to scaling percolation law, which yields the following values of threshold concentrations: $C_n^{p_1}\approx 0.0002$ wt. \% and  $C_n^{p_2}\approx  0.5$ wt. \%. The two step percolation was described by using the mean field formalism for the particles with a core-shell structure demonstrating various electrical conductivities of the cores and shells. The transition from the first to the second percolation can also be revealed for corresponding $\epsilon'(C_n)$ curve if it is presented in a double logarithmic scale. This suggests that both conductivity and polarizability of the composites are sensitive to the second percolation.

\section*{Acknowledgements}
This work was partially funded under the projects 2.16.1.4 (NAS of Ukraine). Authors thank Cheap Tubes for providing samples of short nanotubes involved in this research.

%\bibliography{Tomylko}

%

\end{document}